\documentclass[reprint,aps,prl,showpacs,superscriptaddress,longbibliography]{revtex4-1}
\usepackage[latin9]{inputenc}
\usepackage{amsmath}
\usepackage{amssymb}
\usepackage{graphicx}

\makeatletter

\newcommand*\LyXThinSpace{\,\hspace{0pt}}

\usepackage{epstopdf}
\usepackage{amsfonts}
\usepackage{dcolumn}
\usepackage{bm}
\usepackage{color}

\makeatother

\begin{document}

\title{Few-body perspective of quantum anomaly in two-dimensional Fermi
gases}

\author{X. Y. Yin}

\affiliation{Centre for Quantum and Optical Science, Swinburne University of Technology,
Melbourne, Victoria 3122, Australia}

\author{Hui Hu}

\affiliation{Centre for Quantum and Optical Science, Swinburne University of Technology,
Melbourne, Victoria 3122, Australia}

\author{Xia-Ji Liu}

\affiliation{Centre for Quantum and Optical Science, Swinburne University of Technology,
Melbourne, Victoria 3122, Australia}

\date{\today }
\begin{abstract}
Quantum anomaly manifests itself in the deviation of breathing mode
frequency from the scale invariant value of $2\omega$ in two-dimensional
harmonically trapped Fermi gases, where $\omega$ is the trapping
frequency. Its recent experimental observation with cold-atoms reveals
an unexpected role played by the effective range of interactions,
which requires quantitative theoretical understanding. Here we provide
accurate, benchmark results on quantum anomaly from a few-body perspective.
We consider the breathing mode of a few trapped interacting fermions
in two dimensions up to six particles and present the mode frequency
as a function of scattering length for a wide range of effective range.
We show that the maximum quantum anomaly gradually reduces as effective
range increases while the maximum position shifts towards the weak-coupling
limit. We extrapolate our few-body results to the many-body limit
and find a good agreement with the experimental measurements. Our
results may also be directly applicable to a few-fermion system prepared
in microtraps and optical tweezers. 
\end{abstract}

\maketitle

Quantum anomaly arises if a certain symmetry of a classical theory 
(i.e., action) of the system fails to hold when a full quantum description 
is developed. One important example is the deviation from
the scale invariant breathing mode frequency in two-dimensional (2D)
ultracold atomic quantum gases \cite{pitaevskii1997,olshanii2010}.
In a classic 2D gas interacting through zero-range interaction, a
hidden $SO(2,1)$ symmetry leads to an \emph{exact} breathing mode
frequency of $\omega_{B}=2\omega$, where $\omega$ is the trapping
frequency \cite{pitaevskii1997}. In a 2D quantum gas, the requirement
of renormalization of the contact interaction leads to an additional
length scale, the 2D scattering length. Therefore, the breathing mode
frequency $\omega_{B}$ deviates from $2\omega$ and depends on the
scattering length. This is in contrast to the three-dimensional unitarily
interacting quantum gas, where infinitely strong interaction strength
leads to scale invariance and the preservation of $SO(2,1)$ symmetry
\cite{werner2006}.

Experimental verification of quantum anomaly in 2D quantum gases comes
with several interesting surprises. The initial idea of a weakly-interacting
2D Bose gas \cite{pitaevskii1997,olshanii2010} does not work since
the breathing mode frequency shift is too small to be observable.
A strongly interacting two-component 2D Fermi gas of $^{40}$K and
$^{6}$Li atoms appears to be a better candidate \cite{hofmann2012,taylor2012,murthy2019}.
However, the early observation of about $1\%$ frequency shift from
$2\omega$ \cite{vogt2012} is much smaller than the $10\%$ anomaly
predicted for zero-range contact interaction at zero temperature \cite{hofmann2012,taylor2012}.
This discrepancy is probably caused by the large temperature of the
$^{40}$K Fermi gas in the experiment \cite{vogt2012}, as suggested
by virial expansion studies \cite{chafin2013,mulkerin2018}. In two
most recent experiments with $^{6}$Li atoms, the 2D Fermi gas was
cooled down to one-tenth of Fermi temperature, to avoid any dominant
finite temperature effect \cite{holten2018,peppler2018}. Surprisingly,
the measured quantum anomaly, about $2-3\%$, is still at the same
level as in the first observation \cite{vogt2012}. This unexpected,
much reduced quantum anomaly is now understood as a result of a significant
effective range of interactions induced by the tight axial confinement
\cite{petrov2001,hu2019,wu2019} that is necessary to restrict the
motion of atoms into two dimensions \cite{petrov2001,turlapov2017}.
An approximate many-body theory that takes into account Gaussian pair
fluctuations (GPF) has then been developed, giving rise to a qualitative
explanation for the experimental observation \cite{hu2019}. More
accurate many-body calculations (except quantum Monte Carlo simulations
\cite{shi2015,anderson2015,schonenberg2017}) are difficult to carry
out at finite effective range, due to strong correlation inherent
in 2D interacting Fermi gases.

In this Letter, we aim to provide benchmark predictions on quantum
anomaly from a \emph{few-body} perspective. Our purpose is three-fold.
First, few-body systems can be solved exactly with high accuracy.
Therefore, the few-body results include all correlations that can
not be accounted for in the approximate mean-field and GPF studies.
Here we use explicitly correlated Gaussian (ECG) basis set expansion
approach \cite{cg_book,mitroy2013} to understand the breathing mode frequency
in a small Fermi cloud with up to six particles. Second, few-body
systems provide a bridge between the two-body physics and many-body
physics. Many results in the many-body limit can be successfully deduced
from few-body studies \cite{liu2009,liu2010,blume2012,liu2013,bruun2016,levinsen2017}.
In this work, we find that our few-body results reveal two important
features observed in the experiments in the presence of a nonzero
effective range \cite{holten2018,peppler2018}: the maximum quantum
anomaly reduces gradually with increasing effective range while the
peak position shifts to the weakly-interacting regime. Finally, it
is worth emphasizing that the few-fermion system under consideration
can nowadays be routinely prepared by using microtraps \cite{serwane2011}
or optical tweezer \cite{kaufman2012,liu2018}. Thus, the few-body
result should be of its own interest and we anticipate that our few-body
predictions with tunable number of particles might be directly examined
in future experiments. 

\textit{Model Hamiltonian}. \textemdash{} We start by mentioning the
$s$-wave scattering phase shift of two particles in 2D, which can
be expanded in the low-energy limit as \cite{verhaar1984, adhikari1986} 
\begin{equation}
\cot[\delta(k)]=\frac{2}{\pi}\left[\gamma+\ln\left(\frac{ka_{2D}}{2}\right)\right]+\frac{1}{2}k^{2}R_{s}+\cdots,
\end{equation}
where $\gamma\simeq0.577$ is Euler's constant, $a_{2D}$ the 2D scattering
length and $R_{s}$ the effective range \cite{Note2DScatteringLength}.
$R_{s}$ is related to the asymptotic form of the normalized scattering
wave function $u(r)$ of a particular interaction potential by 
\begin{equation}
R_{s}=2\int_{0}^{\infty}\left[\ln^{2}\left(r/a_{2D}\right)-u^{2}\left(r\right)\right]rdr.
\end{equation}
In 2D, the scattering length $a_{2D}$ is always positive. The two-component
Fermi gas goes through a crossover from a Bose-Einstein condensate
(BEC) to a Bardeen-Cooper-Schrieffer (BCS) superfluid as $a_{2D}$
increases. However, unlike 3D, there is no scale invariant unitarity
limit where the interaction strength diverges. Instead, the strongly
interacting regime is around $\ln(k_{F}a_{2D})=0$, where $k_{F}$
is the characteristic Fermi wavevector.

\begin{figure}[t]
\centering{}\includegraphics[bb=10bp 10bp 497bp 546bp,clip,width=0.5\textwidth]{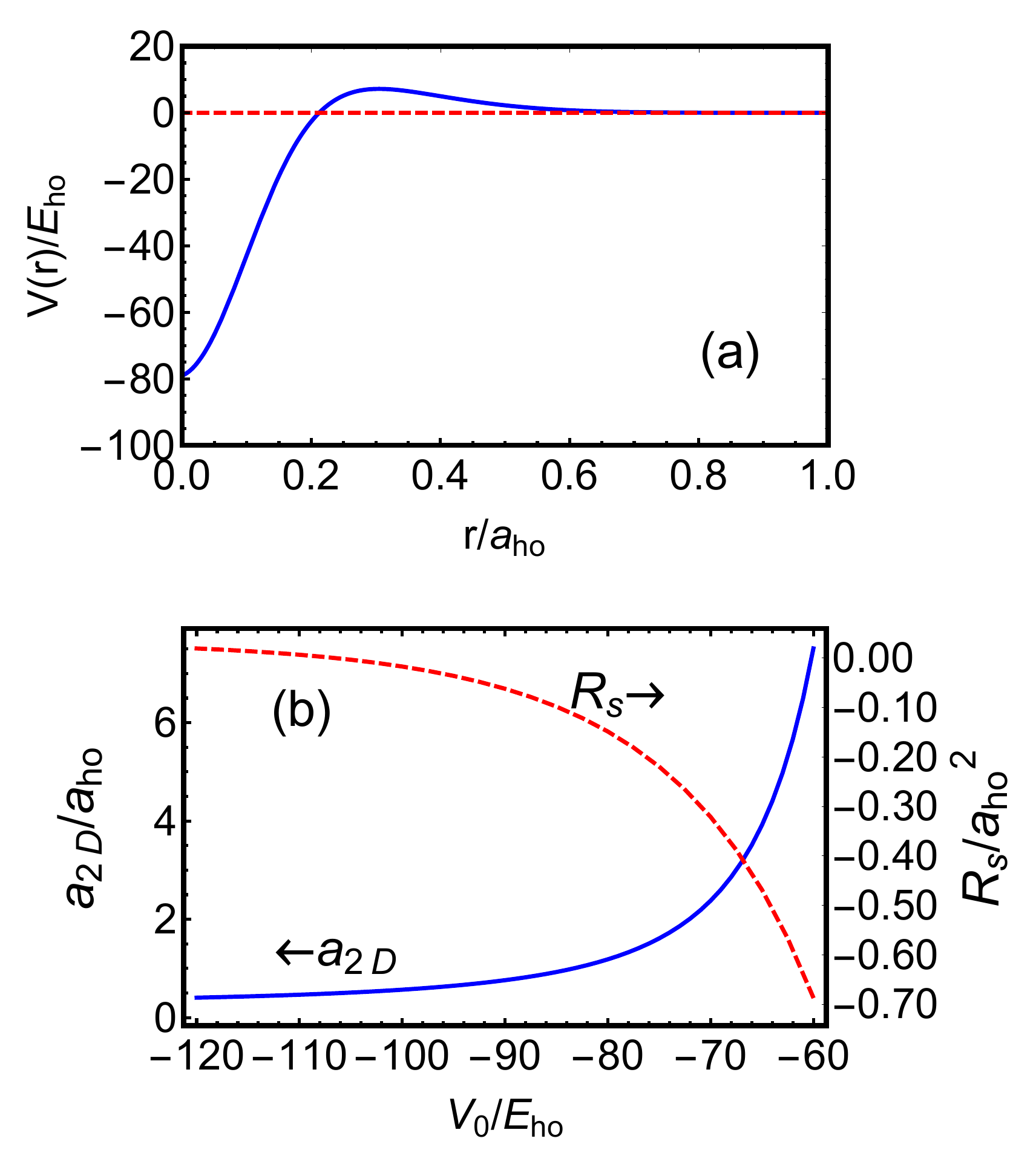}
\caption{(a) Pseudopotential $V(r)$ at $V_{0}/E_{\text{ho}}=-79.04$ and $r_{0}/a_{\text{ho}}=0.104$.
Here, $E_{\text{ho}}=\hbar\omega$ and $a_{\text{ho}}=\sqrt{\hbar/(m\omega)}$
are the energy and length of the harmonic oscillator, respectively.
This potential coincides with the potential used in obtaining the third diamond symbol of Fig.~2(c). 
Dashed line marks the line $V(r)=0$. (b) The $2D$ scattering length
$a_{2D}$ (solid line, scale on the left) and the effective range
$R_{s}$ (dashed line, scale on the right) as a function of potential
depth $V_{0}$ at $r_{0}/a_{\text{ho}}=0.1$.}
\label{fig_potential} 
\end{figure}

We consider a two-component Fermi gas consisting of $N_{\uparrow}$
spin-up and $N_{\downarrow}$ spin-down atoms ($N_{\uparrow}=N_{\downarrow}=N$)
under harmonic confinement with transverse trapping frequency $\omega$.
The system Hamiltonian reads,
\begin{equation}
\mathcal{H}=\sum_{\sigma}\sum_{i_{\sigma}}\left[-\frac{\hbar^{2}\nabla_{i_{\sigma}}^{2}}{2m}+\frac{m\omega^{2}\vec{r}_{i_{\sigma}}^{2}}{2}\right]+\sum_{i_{\uparrow},i_{\downarrow}}V\left(r_{i_{\uparrow}i_{\downarrow}}\right),\label{eq_H}
\end{equation}
where $m$ is the mass of a single atom, $\vec{r}_{i_{\sigma}}$ ($i_{\sigma}=1,\cdots,N$
and $\sigma=\uparrow,\downarrow$) denotes the 2D position vector
of the $i$th spin-up or down atom with respect to the trap center,
and $V(r)$ is the interspecies two-body interaction potential that
depends on the distance $r_{i_{\uparrow}i_{\downarrow}}\equiv|\vec{r}_{i_{\uparrow}}-\vec{r}_{i_{\downarrow}}|$.

The effective range $R_{s}$ induced by a tight axial confinement
in 2D Fermi gas experiments is negative \cite{hu2019}. To simulate
a negative $R_{s}$, we employ a pseudopotential,
\begin{equation}
V(r)=V_{0}\exp\left(-\frac{r^{2}}{2r_{0}^{2}}\right)-V_{0}\frac{r}{a_\text{ho}}\exp\left[-\frac{r^{2}}{2(2r_{0})^{2}}\right],\label{eq_potential}
\end{equation}
which is numerically amenable to ECG simulations. Alternative pseudopotential
is also examined to check the negligible effects beyond effective range~\cite{SM}.
As exemplarily shown in Figure~\ref{fig_potential}(a), the potential exhibits two features
that are essential for supporting a shape resonance: an attractive
well that supports virtual bound states and a potential barrier that
couples the virtual bound state to the free-space scattering states
\cite{schonenberg2017}. Negative $R_{s}$ becomes possible near a
shape resonance. Figure~\ref{fig_potential}(b) shows the 2D scattering
length $a_{2D}$ and effective range $R_{s}$ as a function of the
depth of potential $V_{0}$ for a fixed $r_{0}$. As the depth of
potential increases, both $a_{2D}$ and the absolute value of $R_{s}$
decreases until the next bound state appears. 
Another interesting feature of the potential is that the
volume is positive, and therefore cannot support a two-body bound state
when $V_{0}\rightarrow0$.
Here, we restrict ourself
to potential that supports at most one two-body bound state in free
space. By adjusting both $V_{0}$ and $r_{0}$, one can achieve a
wide range of combination of $a_{2D}$ and $R_{s}$. The limitation
for our potential is that it is not possible to produce a small $a_{2D}$
and large $|R_{s}|$ at the same time.

We solve the time-independent Schrödinger equation for the Hamiltonian
given in Eq.~(\ref{eq_H}) using ECG basis set expansion approach
\cite{cg_book,mitroy2013,SM}. We expand the eigenstates of the Hamiltonian in
terms of ECG basis functions, which depend on a number of nonlinear
variational parameters that are optimized through energy minimization
\cite{mitroy2013,blume2009,blume2011,yin2015}. After a proper basis
set is constructed, we calculate various ground-state properties of
a few-fermion system with number of particles $(N_{\uparrow},N_{\downarrow})=(1,1)$,
$(2,2)$ and $(3,3)$. For a trapped system, it is convenient to use
Fermi wavevector $k_{F}=\sqrt{2}(N_{\text{tot}})^{1/4}a_{\text{ho}}^{-1}$,
which is basically the wavevector of a non-interacting Fermi gas at
the trap center in the large-particle limit \cite{turlapov2017}.
Here, $N_{\text{tot}}\equiv N_{\uparrow}+N_{\downarrow}$ is the total
number of atoms and $a_{\text{ho}}=\sqrt{\hbar/(m\omega)}$ the harmonic
oscillator length.

\begin{figure}[t]
\centering{}\includegraphics[width=0.48\textwidth]{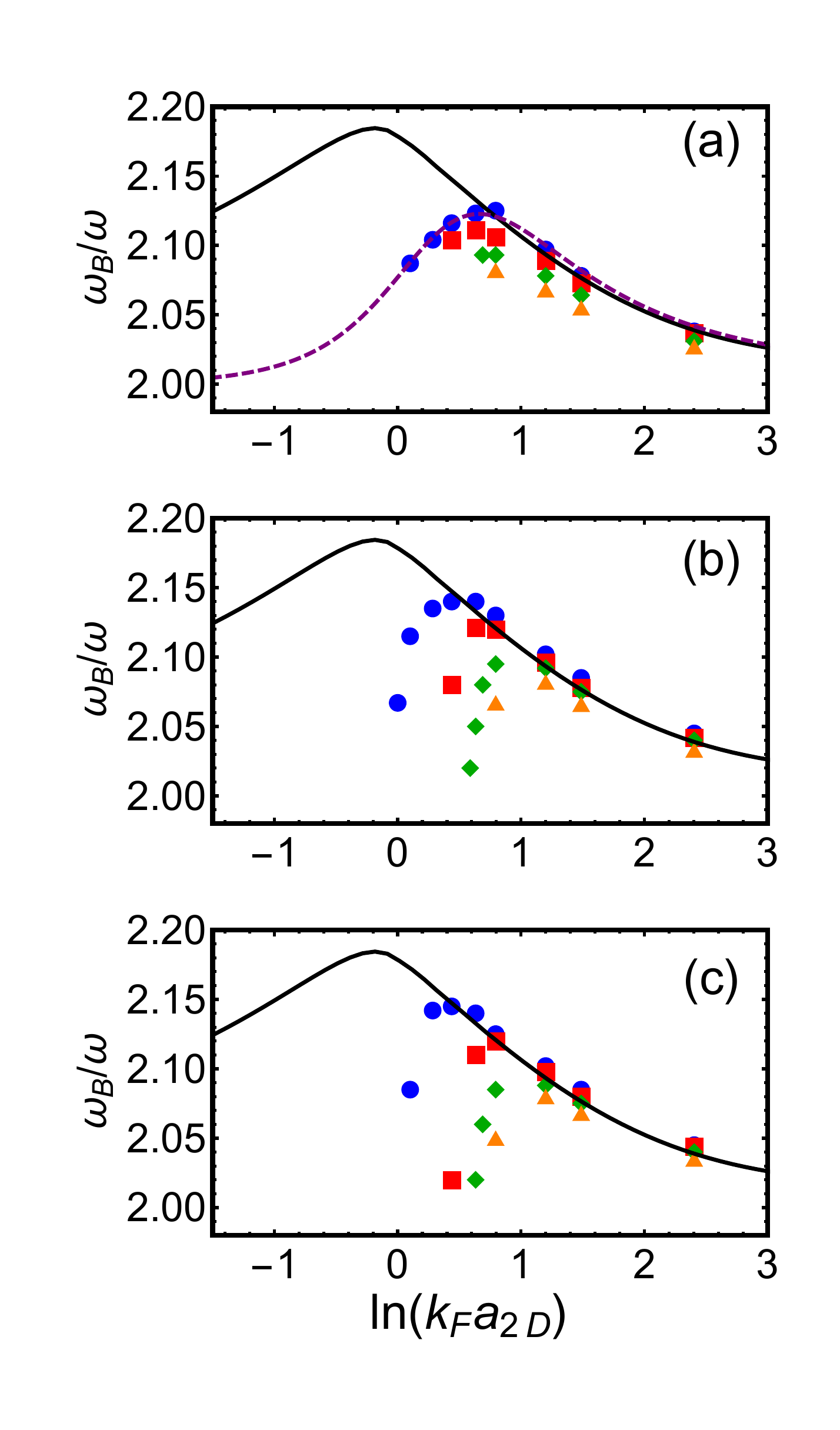}
\caption{Symbols in (a), (b), and (c) show the breathing mode frequency $\omega_{B}$
as a function of logarithm of scattering length for $(1,1)$, $(2,2)$,
and $(3,3)$ systems, respectively. Blue circles, red squares, green
diamonds, and orange triangles are for $k_{F}^{2}R_{s}=-0.0245$,
$-0.245$, $-0.620$, and $-0.980$, respectively. The solid line
shows the zero-range result in the many-body limit based on AFQMC
equation of state while the dashed line shows the exact two-body zero-range solution
\cite{SM}.}
\label{fig_freq} 
\end{figure}

\textit{Breathing mode frequency}. \textemdash{} The breathing mode
of two-component Fermi gases is a density fluctuation excited by the
perturbation $\mathcal{F}\propto\sum\vec{r}_{i\sigma}^{2}$. It appears
as a well-defined single mode in the density response function in
the low-momentum and low-energy limit, i.e., $S_{\mathcal{F}}(\omega)\propto\delta(\omega-\omega_{B})$.
As a result, we may use a sum-rule approach to calculate the breathing
mode frequency, $\omega_{B}^{2}=m_{p+2}/m_{p}$, through the energy
weighted moments $m_{p}\equiv\int d\omega\omega^{p}S_{\mathcal{F}}(\omega)$.
In harmonic traps, both the energy weighted moments $m_{1}$ and $m_{-1}$
can be easily calculated at zero temperature. The former can be transformed
into the calculation of commutators involving the operator $\mathcal{F}$
and the Hamiltonian \cite{dalfovo1999}, leading to $m_{1}\propto\langle r^{2}\rangle=\int_{0}^{\infty}n(r)r^{3}dr$,
where $\langle r^{2}\rangle$ is the square cloud width calculated
using the density distribution $n(r)$. The moment $m_{-1}$, on the
other hand, is proportional to the linear static response of the operator
$\mathcal{F}$, $m_{-1}\propto\partial\langle r^{2}\rangle/\partial(\omega^{2})$
\cite{dalfovo1999}. Putting the two moments together, we obtain an
elegant expression for the breathing mode frequency \cite{menotti2002,hu2014},
\begin{equation}
\omega_{B}^{2}=-2\langle r^{2}\rangle\left[\frac{\partial\langle r^{2}\rangle}{\partial(\omega^{2})}\right]^{-1}.
\end{equation}
The accuracy of the above sum-rule expression may be examined in the
many-particle limit, by using alternative two-fluid hydrodynamic theory
\cite{taylor2009} for the breathing mode frequency in case of zero
range interaction. We outline the details in Supplemental Material
\cite{SM}. For our few-body calculations, which are performed in
harmonic oscillator unit, we define $a_{2D}=xa_{\text{ho}}$, $R_{s}=ya_{\text{ho}}^{2}$,
and $\left<r^{2}\right>=za_{\text{ho}}^{2}$, where $x$, $y$, and
$z$ are dimensionless. If we keep $\omega$ a constant while vary
$a_{2D}$ and $R_{s}$ proportionally, i.e., keep $y/x^{2}$ a constant,
the sum rule result can be alternatively expressed as 
\begin{equation}
\omega_{B}^{2}=4\omega^{2}/\left(1-\frac{x}{2z}\frac{dz}{dx}\right).
\end{equation}
This allows us to compute $\omega_{B}$ for few-body systems through
a finite difference method \cite{SM}. To compare $\omega_{B}$ for
$(1,1)$, $(2,2)$, and $(3,3)$ systems, and to provide insights
for larger systems, we express the harmonic oscillator length $a_{\textrm{ho}}$
in terms of Fermi wavevector $k_{F}$ and consequently use the dimensionless
interaction parameters $\ln(k_{F}a_{2D})$ and $k_{F}^{2}R_{s}$.

Symbols in Fig.~\ref{fig_freq}(a), (b), and (c) show the breathing
mode frequency $\omega_{B}$ as a function of $\ln(k_{F}a_{2D})$
for various effective ranges $k_{F}^{2}R_{s}$, ranging from close
to zero to around $-1$. The solid line shows the zero-range many-body
results determined using the equation of state from auxiliary-field
quantum Monte Carlo (AFQMC) \cite{SM}. In all regimes, $\omega_{B}$
is greater than the scale invariance value $2\omega$. It peaks within
strongly interacting regime, gradually decreases towards the BCS limit,
and decreases sharply towards the BEC limit. 

As the effective range $k_{F}^{2}R_{s}$ approaches zero, our finite-range
few-body results approaches the zero-range many-body result. The difference
between $k_{F}^{2}R_{s}=-0.0245$ results, where $k_{F}^{2}R_{s}$
is very close to zero, and the zero-range many-body result is very
small on the BCS side and could likely be attributed to the finite
size effect. 
However, the discrepancy becomes prominent when $\ln(k_{F}a_{2D})<0.2$.
This may be caused by the dramatically increased peak density in the
strongly interacting regime, where the wavevector $k_{F}$ of the
non-interacting trapped Fermi gas is no longer a proper choice for
parameterization. The actual effective range at the trap center is
larger, leading to a significant deviation in the breathing mode frequency
with respect to the zero-range result. 
Notably, our numerical result for $(1,1)$ system with $k_{F}^{2}R_{s}=-0.0245$ 
agrees extremely well with the prediction from the exact two-body zero-range 
solution~\cite{SM}, based on the seminal work of Busch et al.~\cite{busch1998}.

As $k_{F}^{2}R_{s}$ increases, the quantum anomaly, i.e., the difference
between $\omega_{B}$ and scale invariance value $2\omega$, is gradually
suppressed across all scattering strength. Specifically, as $k_{F}^{2}R_{s}$
increases from near zero to around $-1$, the maximum value of breathing
mode shift decreases from $7.5\%$ to $4\%$. Also, the scattering
length where maximum value of $\omega_{B}$ occurs shifts towards
the BCS limit as $k_{F}^{2}R_{s}$ increases. The downward shifting
of $\omega_{B}$ as the effective range increases qualitatively agrees
with the trend observed in the many-body calculations \cite{hu2019,wu2019}.

\begin{figure}[t]
\centering{}\includegraphics[bb=15bp 15bp 540bp 652bp,width=0.5\textwidth]{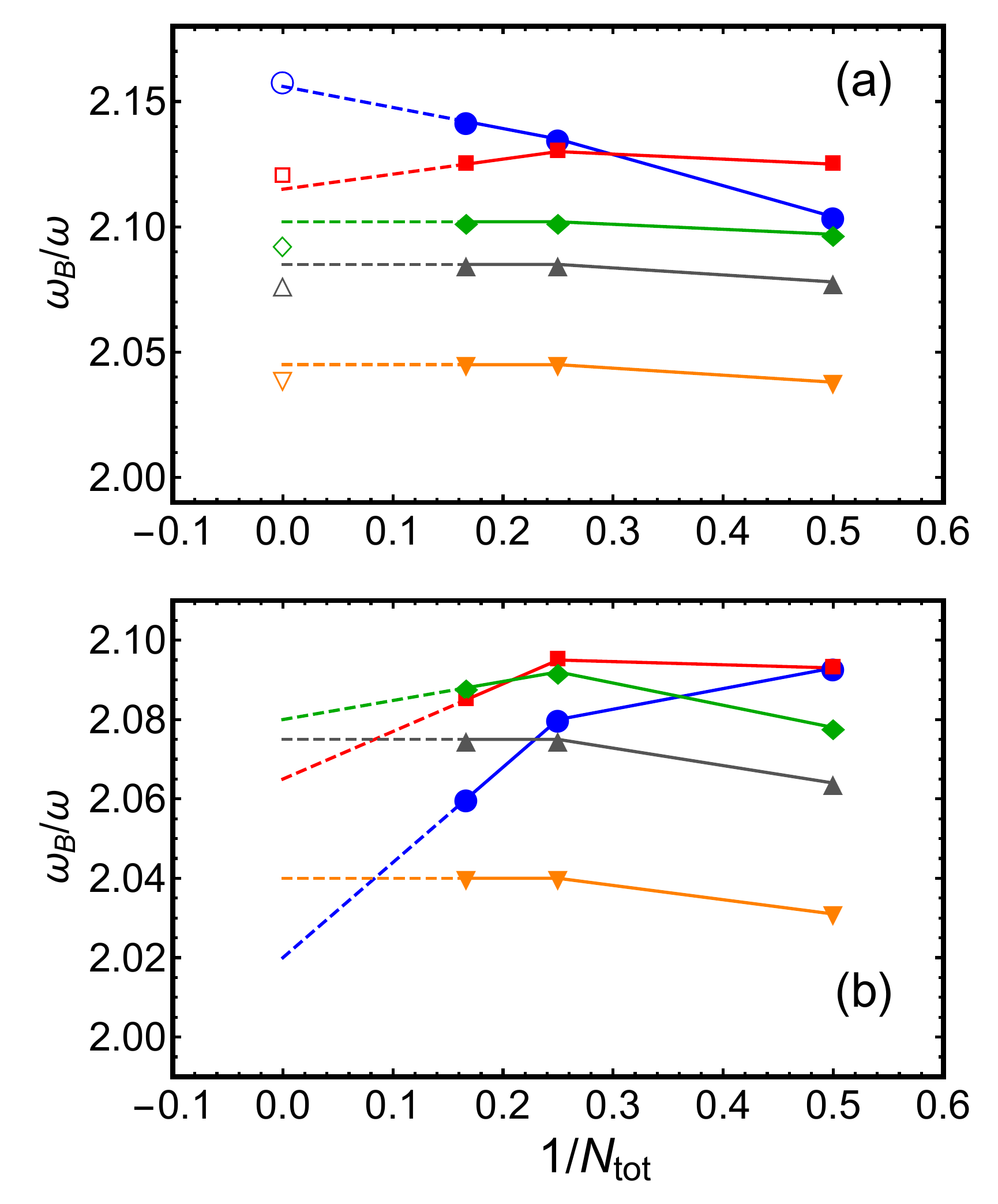}
\caption{(a) Circles, squares, diamonds, upper triangles, and lower triangles
show the breathing mode frequency $\omega_{B}$ as a function of $1/N_{\text{tot}}$
at $\ln(k_{F}a_{2D})=0.284$, $0.794$, $1.20$, $1.49$, and $2.40$,
respectively. The effective range is $k_{F}^{2}R_{s}=-0.0245$. The
hollow symbols at $1/N_{\text{tot}}=0$ show the many-body result
at zero effective range. (b) Circles, squares, diamonds, upper triangles,
and lower triangles show the breathing mode frequency $\omega_{B}$
as a function of $1/N_{\text{tot}}$ at $\ln(k_{F}a_{2D})=0.689$,
$0.794$, $1.20$, $1.49$, and $2.40$, respectively. The effective
range is $k_{F}^{2}R_{s}=-0.620$. In both plots, dashed lines show
the linear extrapolation of $(2,2)$ and $(3,3)$ results towards
the many-particle limit $N_{\text{tot}}\rightarrow\infty$.}
\label{fig_conv} 
\end{figure}

Although we consider only $(1,1)$, $(2,2)$, and $(3,3)$ systems,
it is important to analyze the finite size effects and the trend as
the system size increases. This will allow us to see how our few-body
results connect to the many-body limit. Solid symbols in Fig.~\ref{fig_conv}(a)
show $\omega_{B}$ as a function of $1/N_{\text{tot}}$ for $k_{F}^{2}R_{s}=-0.0245$
and different values of $\ln(k_{F}a_{2D})$ across the BEC-BCS crossover.
Overall, the differences between the $(2,2)$ and $(3,3)$ systems
are very small. Since $(1,1)$ system cannot capture the many-body
correlations beyond two-body level, it is expected to see a significant
difference between $(1,1)$ system and larger systems. If we do a
simple linear extrapolation of $(2,2)$ and $(3,3)$ results towards
the $1/N_{\text{tot}}=0$ limit~\cite{NoteExtrapolation}, the extrapolated results agree reasonably
well with the many-body AFQMC result shown in hollow symbols. The
good agreement encourages us to perform the same extrapolation for
the data with effective range $k_{F}^{2}R_{s}=-0.620$, which is the
realistic effective range in recent two $^{6}$Li experiments \cite{holten2018,peppler2018,hu2019}.
This is shown in Fig.~\ref{fig_conv}(b) by dashed lines. From the
results at zero-range, we estimate that the systematic error in the
extrapolated mode frequency due to our naïve extrapolation procedure
is about $(0.01-0.02)\omega$.

\begin{figure}[t]
\centering{}\includegraphics[width=0.45\textwidth]{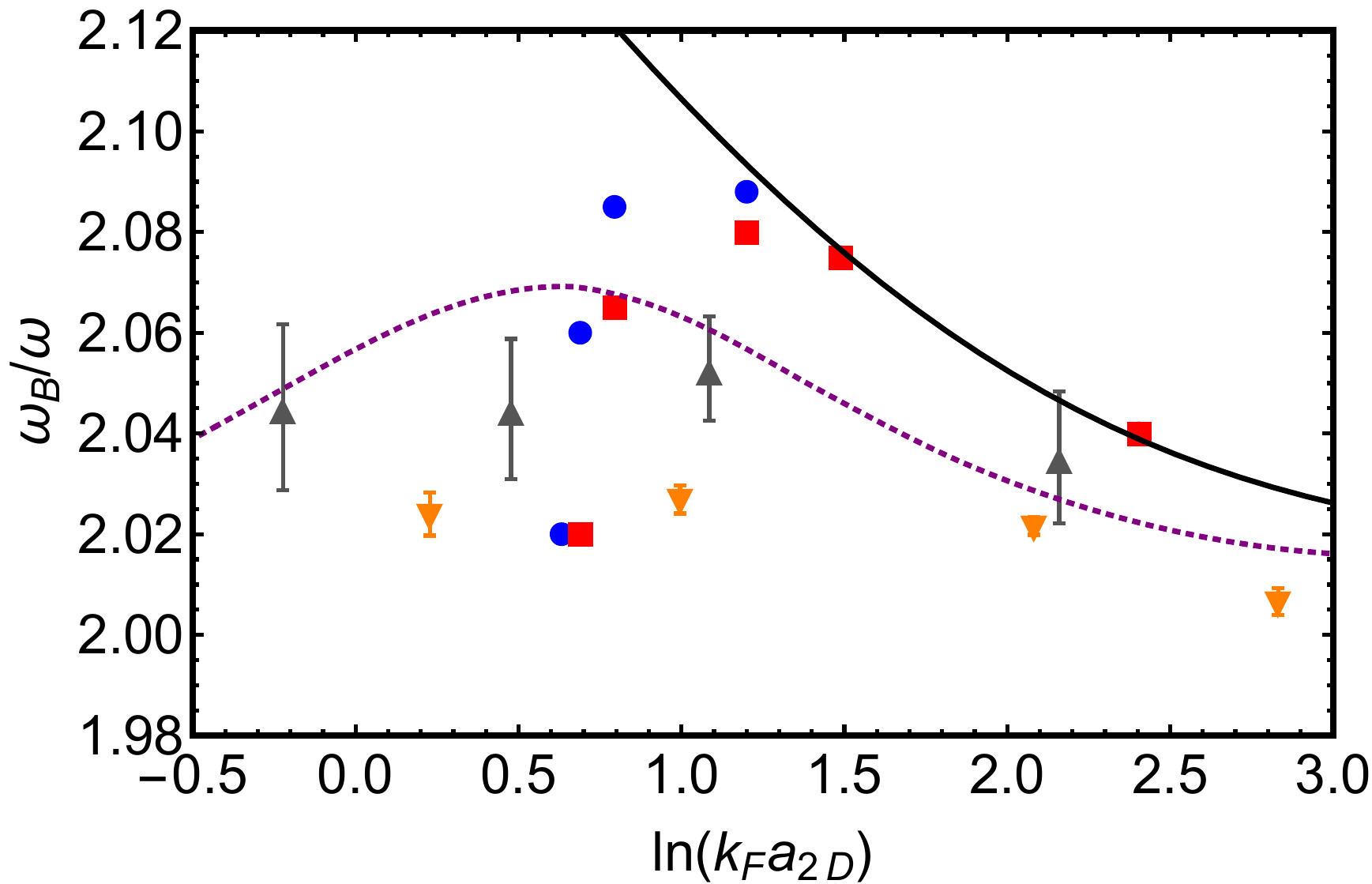} 
\caption{Circles and squares show the breathing mode frequency $\omega_{B}$
as a function of $\ln(k_{F}a_{2D})$ for $(3,3)$ system and for the
extrapolated many-body result, respectively. The effective range is
fixed to $k_{F}^{2}R_{s}=-0.620$. Upper and lower triangles show
the experimental data by Peppler $et$ $al.$ (Swinburne) \cite{peppler2018}
and Holten $et$ $al.$ (Heidelberg) \cite{holten2018}. The solid
line shows the zero-range AFQMC result in the many-body limit
while the dotted line shows the latest many-body result~\cite{wu2019}.}
\label{fig_exp} 
\end{figure}

\textit{Comparison with experiments}. \textemdash{} The main motivation
for this work is the discrepancy between the large quantum anomaly
($\sim10\%$) predicted with zero-range model and the much smaller
quantum anomaly ($<3\%$) observed in the experiments \cite{holten2018,peppler2018}.
In both Heidelberg and Swinburne experiments, two hyperfine states
of $^{6}$Li are cooled to deep quantum degeneracy in a 2D trap. It
has been shown that the effective range in those setups are non-negligible
\cite{hu2019}. Here, we compare the experimental data with our few-body
results with same value of $k_{F}^{2}R_{s}=-0.620$ in Fig.~\ref{fig_exp}.
Circles and squares show the $(3,3)$ system results and extrapolated
many-body results following Fig.~\ref{fig_conv}(b). Upper and lower
triangles show the experimental data from Swinburne and Heidelberg
groups \cite{holten2018,peppler2018}, respectively. 

Although we have observed a reduced quantum anomaly with finite effective
range, quantitative discrepancy between experimental data and our
few-body results exist. One of the reasons could be the residual finite
temperature effect. The temperature in both experiments is at the
range of $(0.1-0.2)T_{F}$, where $T_{F}$ is Fermi temperature. It
has been shown that the quantum anomaly is reduced at non-zero temperature
\cite{mulkerin2018}. Combining the finite range effect and the finite
temperature effect, we expect that the quantum anomaly should be further
reduced. These two effects together, could possibly explains the much
smaller deviation observed in the experiments \cite{holten2018,peppler2018}.

\textit{Conclusions}. \textemdash{} We have calculated the breathing
mode frequency of a two-component Fermi cloud with a few particles
up to six in two dimensions. We have found that the quantum anomaly,
i.e., the enhancement of the breathing mode frequency with respect
to its classical scale invariant value, is gradually suppressed as
effective range of interactions increases. In particular, the maximum
value of the breathing mode frequency is reduced while the interaction
strength corresponding to the maximum value shifts towards the weak-coupling
limit. These two key trends qualitatively agree with recent experiments.
Our accurate few-body results open a new way to better understand
the intriguing quantum anomaly and provide benchmark predictions for
future few-particle experiments in microtraps and optical tweezers.
\begin{acknowledgments}
We are grateful for discussions with Jia Wang. This research was supported
by the Australian Research Council (ARC) Discovery Programs, Grants
No. DP170104008 (H.H.), No. FT140100003 (X.-J.L) and No. DP180102018
(X.-J.L).
\end{acknowledgments}


\begin{thebibliography}{10}
\bibitem{Note2DScatteringLength}Note that the definition of scattering
length and effective range in 2D varies in literature. In our choice,
$R_{s}$ has dimension of $length^{2}$.
\end{thebibliography}
\end{document}

\title{Supplemental Material for ``Few-body perspective of quantum anomaly in two-dimensional Fermi
gases''}

\author{X. Y. Yin}

\affiliation{Centre for Quantum and Optical Science, Swinburne University of Technology,
Melbourne, Victoria 3122, Australia}

\author{Hui Hu}

\affiliation{Centre for Quantum and Optical Science, Swinburne University of Technology,
Melbourne, Victoria 3122, Australia}

\author{Xia-Ji Liu}

\affiliation{Centre for Quantum and Optical Science, Swinburne University of Technology,
Melbourne, Victoria 3122, Australia}

\maketitle

\section{Explicitly correlated Gaussian technique and finite difference method}
To numerically solve the Schr\"odinger equation for the
Hamiltonian given in Eq.~(3), we employ explicitly correlated Gaussian (ECG)
technique detailed in Ref.~\cite{mitroy2013, cg_book}.
First, we separate off the center of mass degrees of freedom and
expand the eigenstates of the relative Hamiltonian
in terms of ECG basis functions, which depend on a set of non-linear
variational parameters that are optimized through
energy minimization. 
Because the interaction we use is rotational invariant,
different angular momentum channels with different $m$ quantum number
do not couple, we focus on $m=0$ channel here.
The unsymmetrized basis functions for states with
$m=0$ quantum number
read $\exp(-\frac{1}{2}\vec{x}^{T}A\vec{x})$,
where $A$ is a symmetric and positive definite
$(N-1)\times(N-1)$ parameter matrix.
$\vec{x}=(\vec{x}_1, \vec{x}_2, ..., \vec{x}_{N-1})^T$ 
collectively denotes a set of
$N-1$ Jacobi vectors.
We choose the Jacobi vectors in the same way as detailed in Ref.~\cite{yin2015}.
A key advantage of these basis functions is that the
corresponding overlap and Hamiltonian matrix
elements can be calculated analytically~\cite{cg_book}.
The fermionic exchange symmetry is ensured by acting
with the permutation operator on the
unsymmetrized basis functions. 

We use a stochastic 
variational approach to choose
and optimize the variational parameters contained in
$A$. We start with just one basis function and enlarge the basis set
one basis function at a time, which is
chosen from 500 trial basis functions.
To decide which trial basis function to keep, we calculate the energy for each of 
the enlarged trial basis sets and choose the one that results in the largest reduction of ground state energy.
The procedure is repeated till the basis set is sufficiently complete.
After the basis set enlargement process, we then readjust each basis function
to further lower the energy. A reoptimization cycle denotes the adjustment process
of every basis functions. Typically, 5 complete reoptimization cycles is needed to reach desired accuracy
for ground state energy.

As we explained in the main text, breathing mode frequency $\omega_B$ can be calculated through Eq.~(6).
To compute $\omega_{B}$ through finite difference method, we choose
two sets of parameters $(x_{1},y_{1})$ and $(x_{2},y_{2})$ that
satisfy $x_{2}=1.02x_{1}$ and $y_{2}=1.02^{2}y_{1}$. Each set of
parameter corresponds to a potential with a certain $V_{0}$ and $r_{0}$.
Since the parameters for the two potentials are extremely close, we
average the two sets of $V_{0}$ and $r_{0}$ and construct the basis
set according to the averaged potential through the aforementioned basis set enlargement
process and 5 complete reoptimization cycles. 
We then compute $z_{1}$
and $z_{2}$ using the same basis set. The breathing mode frequency
is given by 
\begin{equation}
\omega_{B}^{2}=4\omega^{2}/\left(1-\frac{1}{2}\frac{z_{2}/z_{1}-1}{x_{2}/x_{1}-1}\right).
\end{equation}
For $(1,1)$ and $(2,2)$ systems, we alternatively compute $z_{1}$
and $z_{2}$ by constructing two basis sets separately and find good
agreement with using the same basis set. Our approach to compute $z_{1}$
and $z_{2}$ using the same basis set allows for a much faster convergence
of $\omega_{B}$, especially for the $(3,3)$ systems.

While the varational principle guarantees that the ground state energy obtained is a variational upper bound.
The same is not true for $\omega_B$. It is therefore much more challenging to estimate
the error bar for $\omega_B$. Following the spirit of Ref.~\cite{yin2015},
we estimate the error bar from the change of $\omega_B$ within last 3 reoptimization cycles.
Under this assumption, the error bar for $(1,1)$, $(2,2)$, and $(3,3)$ systems are estimated 
to be $\Delta \omega_B/\omega=\pm0.0005$, $\pm0.002$, and $\pm0.008$, respectively.

\section{Breathing mode frequency from the two-fluid hydrodynamic theory}

In the case of zero-range interaction, we have the accurate zero-temperature
equation of state provided by auxiliary-field quantum Monte Carlo
(AFQMC) \cite{shi2015}. In the presence of the harmonic trap $V_{ext}(r)=m\omega^{2}r^{2}/2$,
the density distribution $n(r)$ can then be determined using the
standard local density approximation (LDA) \cite{dalfovo1999}. In
addition to the sum-rule approach, we can use the coupled two-fluid
hydrodynamic equations to calculate the breathing mode frequency \cite{taylor2009}.
The normal modes of those equations at frequency $\Omega$ are most
easily obtained by minimizing a variational action, which, in terms
of the displacement field $\mathbf{u}(\vec{r})$, takes the following
form \cite{taylor2009},

\begin{eqnarray}
\mathcal{S} & = & \frac{1}{2}\int d\vec{r}\left[m\Omega^{2}n\mathbf{u}^{2}+2n\left(\mathbf{\nabla}\cdot\mathbf{u}\right)\left(\mathbf{\nabla}V_{ext}\cdot\mathbf{u}\right)+\right.\nonumber \\
 &  & \left.\left(\mathbf{\nabla}n\cdot\mathbf{u}\right)\left(\mathbf{\nabla}V_{ext}\cdot\mathbf{u}\right)-n\left(\frac{\partial P}{\partial n}\right)\left(\mathbf{\nabla}\cdot\mathbf{u}\right)^{2}\right],\label{eq:uaAction}
\end{eqnarray}
where $[\partial P/\partial n](r)$ is the derivative of the local
pressure at position $\vec{r}$ with respect to the local density.
For compressional modes (including the lowest breathing mode of interest),
we assume the following polynomial ansatz for the displacement field:
\begin{equation}
\mathbf{u}(r)=\vec{r}\sum_{i=0}^{N_{p}-1}A_{i}r^{i},
\end{equation}
where $\{A_{i}\}$ ($i=0,\cdots,N_{p}-1$) are the $N_{p}$ variational
parameters. By inserting this variational ansatz into the action Eq.
(\ref{eq:uaAction}), we express the action $\mathcal{S}$ as functions
of the $N_{p}$ variational parameters $\{A_{i}\}$. The mode frequencies
$\Omega$ of the $N_{p}$ compressional modes can then be obtained
by minimizing the action $\mathcal{S}$ with respect to the $N_{p}$
parameters. The accuracy of our variational calculations for the breathing
mode can be improved by increasing the value of $N_{p}$ and the calculations
indeed converge very rapidly as $N_{p}$ increases.

In more detail, it is straightforward to show that, the expression
of the action can be written in a compact form, 
\begin{equation}
\mathcal{S}=\frac{1}{2}\Lambda^{\dagger}\mathcal{S}\left(\Omega\right)\Lambda,
\end{equation}
where $\Lambda\equiv\left[A_{0},\cdots,A_{i},\cdots,A_{N_{p}-1}\right]^{T}$
and $\mathcal{S}(\Omega)$ is a $N_{p}\times N_{p}$ matrix with elements
($i,j=0,\cdots,N_{p}-1$), 
\begin{equation}
\left[\mathcal{S}\left(\Omega\right)\right]_{ij}\equiv M_{ij}\Omega^{2}-K_{ij},\label{eq:matrixsw}
\end{equation}
where the weighted mass moments are, 
\begin{equation}
M_{ij}=m\int d\vec{r}r^{i+j+2}n\left(r\right),\label{eq:MassMoment}
\end{equation}
and the spring constants are, 
\begin{equation}
K_{ij}=\left(i+2\right)\left(j+2\right)\int d\vec{r}r^{i+j}\left(n\frac{\partial P}{\partial n}\right)\left(r\right).\label{eq:SpringConstant}
\end{equation}
The minimization of the action $\mathcal{S}$ (which is now in a matrix
form) is equivalent to solving $\mathcal{S}\left(\Omega\right)\Lambda=0$.

\begin{figure}[t]
\centering{}\includegraphics[width=0.45\textwidth]{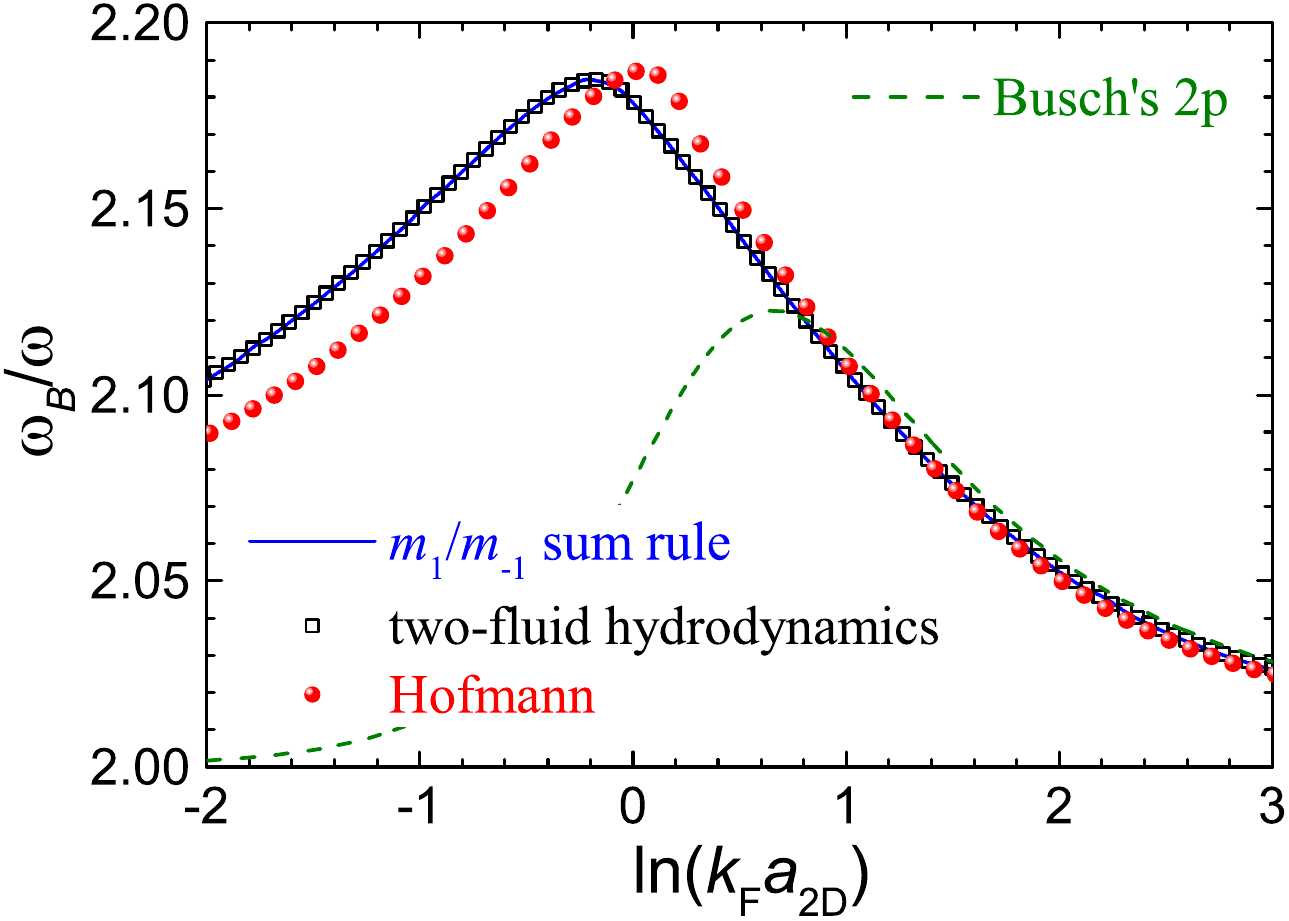} \caption{Breathing mode frequency of a zero-range interacting Fermi gas as
a function of $\ln(k_{F}a_{2D})$, predicted by the sum-rule approach
(solid line), Busch's exact two-body solution (dashed line),
the two-fluid hydrodynamic theory (open squares), and
the simple evaluation adopted by Hofmann (circles) \cite{hofmann2012}.
In all cases, the equation of state simulated by AFQMC is used \cite{shi2015}.
The difference between the sum-rule approach and the two-fluid hydrodynamic
theory is indistinguishable under the scale of the figure.}
\label{fig_zerorangewb} 
\end{figure}

In Fig. \ref{fig_zerorangewb}, we show the breathing mode frequency
predicted by the two-fluid hydrodynamic theory with $N_{p}=10$ (empty
squares), and compare it with the sum-rule result (solid line). The
relative difference between the two-fluid hydrodynamic theory and
the sum-rule approach is less than $0.1\%$ and cannot be identified
at the scale of the figure.

It is interesting to note that, if we assume a \emph{constant} polytropic
index 
\begin{equation}
\gamma=\frac{n}{P}\left(\frac{\partial P}{\partial n}\right)-1,\label{eq:PolytropicIndex}
\end{equation}
the variational procedure leads to a breathing mode frequency 
\begin{equation}
\omega_{B}^{2}=2\left(\gamma+1\right)\omega^{2}.
\end{equation}
This can be seen by substituting the relation $n(\partial P/\partial n)=(\gamma+1)P$
into the spring constants Eq. (\ref{eq:SpringConstant}). Using the
force balance condition in harmonic traps, i.e., $\partial P/\partial r=-m\omega^{2}rn(r)$,
we obtain 
\begin{equation}
K_{ij}=M_{ij}\omega^{2}\frac{\left(i+2\right)\left(j+2\right)}{\left(i+j+2\right)}\left(\gamma+1\right).
\end{equation}
The breathing mode is then decoupled from the other compressional
modes, since for either $i=0$ or $j=0$, $K_{ij}=2(\gamma+1)\omega^{2}M_{ij}$.
This also gives the mode frequency $\omega_{B}=\sqrt{2(\gamma+1)}\omega$.
The assumption of a constant polytropic index was used earlier by
Hofmann to calculate the quantum anomaly with a zero-range contact
interaction \cite{hofmann2012}. In Fig. \ref{fig_zerorangewb}, we
show his results in circles with the updated AFQMC equation of state.
In the weak-coupling limit, this simple estimation agrees very well
with more accurate predictions from the two-fluid hydrodynamic theory
and the sum-rule approach. In the strongly-interacting regime near
$\ln(k_{F}a_{2D})=0$, however, there is a small quantitative difference.

\section{Breathing mode frequency from Busch's exact two-body solution}

For two particles interacting through a zero-range contact potential,
we may also calculate the breathing mode frequency using Busch's exact
two-body solution \cite{busch1998}, which states that the relative
wave-function of the two particles (for $\vec{\rho}=\vec{r}_{1}-\vec{r}_{2}$)
is given by \cite{busch1998,liu2010},
\begin{equation}
\psi_{\textrm{rel}}\left(\rho\right)\propto\exp\left(-\frac{\rho^{2}}{4a_{\textrm{ho}}^{2}}\right)\Gamma\left(-\nu\right)U\left(-\nu,1,\frac{\rho^{2}}{2a_{\textrm{ho}}^{2}}\right),
\end{equation}
where $\Gamma$ is the gamma function and $U$ is the second kind
of Kummer function. The parameter $\nu$ is related to the relative
energy $E_{\textrm{rel}}=(2\nu+1)\hbar\omega$, and is determined
by 
\begin{equation}
\gamma-\frac{1}{2}\ln2+\frac{1}{2}\psi\left(-\nu\right)=\ln\left(\frac{a_{\textrm{ho}}}{a_{2D}}\right),
\end{equation}
where $\gamma\simeq0.577216$ is the Euler constant and $\psi(x)$
is the digamma function. For the ground state, we have $\nu<0$. Adding
the ground-state center-of-mass wave-function (for $\vec{R}=(\vec{r}_{1}+\vec{r}_{2})/2$),
\begin{equation}
\psi_{\textrm{cm}}\left(R\right)\propto\exp\left(-\frac{R^{2}}{a_{\textrm{ho}}^{2}}\right),
\end{equation}
we therefore obtain that the total ground-state wave-function, 
\begin{equation}
\Psi_{\textrm{gs}}\left(\vec{R},\vec{\rho}\right)\propto e^{-\frac{R^{2}}{a_{\textrm{ho}}^{2}}}e^{-\frac{\rho^{2}}{4a_{\textrm{ho}}^{2}}}\Gamma\left(-\nu\right)U\left(-\nu,1,\frac{\rho^{2}}{2a_{\textrm{ho}}^{2}}\right).
\end{equation}
It is now straightforward to calculate the square cloud width, which
takes the form, 
\begin{equation}
\langle r^{2}\rangle=\frac{\int d\vec{R}d\vec{\rho}\left(R^{2}+\rho^{2}/4\right)\left[\Psi_{\textrm{gs}}\left(\vec{R},\vec{\rho}\right)\right]^{2}}{\int d\vec{R}d\vec{\rho}\left[\Psi_{\textrm{gs}}\left(\vec{R},\vec{\rho}\right)\right]^{2}}.
\end{equation}
Here we have used the identity $(r_{1}^{2}+r_{2}^{2})/2=R^{2}+\rho^{2}/4$.
By using the dimensionless variables $y=R^{2}/(2a_{\textrm{ho}}^{2})$
and $x=\rho^{2}/(2a_{\textrm{ho}}^{2})$, we may write $\left\langle r^{2}\right\rangle $
into the form,
\begin{equation}
\langle r^{2}\rangle=\frac{a_{\textrm{ho}}^{2}}{2}\mathcal{F}\left(\nu\right),
\end{equation}
where the dimensionless function is defined by,
\[
\mathcal{F}\left(\nu\right)=\frac{\int dydx\left(4y+x\right)e^{-4y}e^{-x}\Gamma^{2}\left(-\nu\right)U^{2}\left(-\nu,1,x\right)}{\int dydxe^{-4y}e^{-x}\Gamma^{2}\left(-\nu\right)U^{2}\left(-\nu,1,x\right)}.
\]
The integration over the variable $y$ can be easily done and we obtain
\[
\mathcal{F}\left(\nu\right)=\frac{\int dx\left(1+x\right)e^{-x}\Gamma^{2}\left(-\nu\right)U^{2}\left(-\nu,1,x\right)}{\int dxe^{-x}\Gamma^{2}\left(-\nu\right)U^{2}\left(-\nu,1,x\right)}.
\]
We then write $\mathcal{F}(\nu)=1+B_{\nu}/A_{\nu}$, where 
\begin{eqnarray*}
A_{\nu} & = & \int dxe^{-x}\Gamma^{2}\left(-\nu\right)U^{2}\left(-\nu,1,x\right),\\
B_{\nu} & = & \int dxxe^{-x}\Gamma^{2}\left(-\nu\right)U^{2}\left(-\nu,1,x\right).
\end{eqnarray*}
To finish the integral on the variable $x$, we can use the identity
\begin{equation}
\Gamma\left(-\nu\right)U\left(-\nu,1,x\right)=\sum_{k=0}^{\infty}\frac{L_{k}\left(x\right)}{k-\nu}
\end{equation}
to obtain
\begin{eqnarray*}
A_{\nu} & = & \sum_{n=0}^{\infty}\frac{1}{\left(n-\nu\right)^{2}},\\
B_{\nu} & = & \sum_{n=0}^{\infty}\frac{1}{n-\nu}\left[\frac{1+2\nu}{n-\nu}-\frac{1+\nu}{n-\nu-1}-\frac{\nu}{n-\nu+1}\right].
\end{eqnarray*}
Here, $L_{k}(x)$ is Laguerre polynomial, satisfying,
\[
\int_{0}^{\infty}dxe^{-x}L_{m}\left(x\right)L_{n}\left(x\right)=\delta_{mn},
\]
and 
\begin{eqnarray*}
\left(k+1\right)L_{k+1}\left(x\right) & = & \left(2k+1-x\right)L_{k}\left(x\right)-kL_{k-1}\left(x\right).
\end{eqnarray*}
We can now use the $f$-sum rule 

\begin{equation}
\frac{\omega_{B}^{2}}{\omega^{2}}=-2\frac{\langle r^{2}\rangle}{\omega^{2}\left[\frac{\partial\langle r^{2}\rangle}{\partial(\omega^{2})}\right]}
\end{equation}
to calculate the breathing mode frequency. By noting that $a_{\textrm{ho}}^{2}\propto\omega^{-1}$
and $a_{\textrm{ho}}/a_{2D}\propto(\omega^{2})^{-1/4}$, it is easy
to show that,
\begin{equation}
\frac{\omega_{B}^{2}}{\omega^{2}}=4\left[1+\frac{1}{2}\frac{\partial\ln\mathcal{F}}{\partial\ln\left(a_{\textrm{ho}}/a_{2D}\right)}\right]^{-1}.
\end{equation}
For two particles, the Fermi wavevector $k_{F}$ is given by, $k_{F}=2^{3/4}a_{\textrm{ho}}^{-1}$.
Therefore, we have the relation,
\begin{equation}
\ln\left(k_{F}a_{2D}\right)=\frac{3}{4}\ln2-\ln\frac{a_{\textrm{ho}}}{a_{2D}}.
\end{equation}
We can further cast the breathing mode frequency into the form,
\begin{equation}
\omega_{B}=2\omega\left[1-\frac{1}{2}\frac{\partial\ln\mathcal{F}}{\partial\ln\left(k_{F}a_{2D}\right)}\right]^{-1/2}.\label{eq:wb2p}
\end{equation}
Numerically, for a given $\ln(k_{F}a_{2D})$, we first calculate $\nu$
by using 
\begin{equation}
\gamma+\frac{1}{2}\psi\left(-\nu\right)=\frac{5}{4}\ln2-\ln\left(k_{F}a_{2D}\right),
\end{equation}
and then calculate $A_{\nu}$ and $B_{\nu}$ to obtain $\mathcal{F=}(1+B_{\nu}/A_{\nu})$.
This leads to a curve of $\ln\mathcal{F}$ as a function of $\ln(k_{F}a_{2D})$.
We finally take the numerical derivative in Eq. (\ref{eq:wb2p}) to
obtain the breathing mode frequency.

As shown in Fig. S1, we find that the breathing mode frequency from
Busch's exact two-body solution agrees well with the many-body prediction
at $\ln(k_{F}a_{2D})\gtrsim0.7$. Once the interaction parameter is
smaller than this value, there is a significant difference, as the
many-body correlation starts to make a dominant contribution to the
breathing mode. In the main text, we show that the exact two-body solution
agrees very well with the numerical results for $(1,1)$ system with the smallest effective range.

\section{Universality and effects beyond effective range}
In the main text, we discussed how breathing mode frequency changes with regard
to the scattering length $a_{2D}$ and effective range $R_s$. Here, we analyze the
effects beyond the effective range contribution.
In addition to the potential, Eq.~(4), considered in the main text, we additionally consider
a different potential
\begin{equation}
V_s(r)=V_{0}\exp\left(-\frac{r^{2}}{2r_{0}^{2}}\right)-V_{0}\frac{r}{a_\text{ho}}\exp\left[-\frac{r^{2}}{(2r_{0})^{2}}\right].\label{eq_potential2}
\end{equation}
Note that, the potential $V(r)$, Eq.~(4), can produce a much larger range of $a_{2D}$ and $R_s$ for negative
$R_s$ than $V_s(r)$. Therefore, we are only able to compare the results from the two
different potentials for limited data points.

\begin{figure}[t]
\centering{}\includegraphics[width=0.52\textwidth]{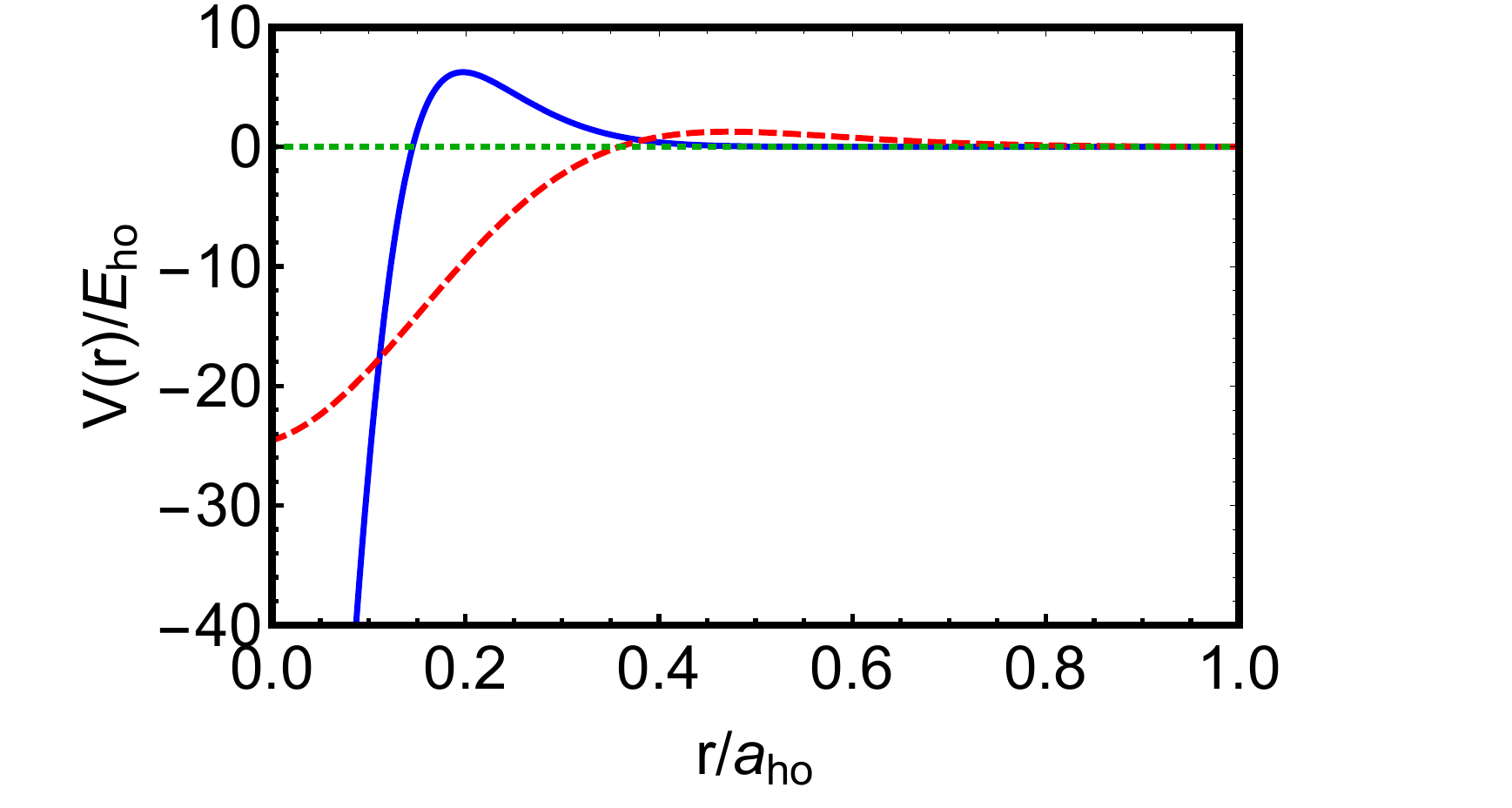}
\caption{Solid line shows pseudopotential $V(r)$ at $V_{0}/E_{\text{ho}}=-120.934$ and $r_{0}/a_{\text{ho}}=0.0641$.
Dashed line shows pseudopotential $V_s(r)$ at $V_{0}/E_{\text{ho}}=-27.7817$ and $r_{0}/a_{\text{ho}}=0.1776$.
Dotted line marks the line $V(r)=0$.}
\label{fig_uni} 
\end{figure}

First, we adjust $V_0$ and $r_0$ in both potentials and plot them in Fig.~\ref{fig_uni}, 
such that the scattering length and effective range
are $a_{2D}=2.213a_\text{ho}$ and $R_s=-0.0612a_\text{ho}^2$, respectively. For $(2,2)$ system,
this corresponds to $k_F a_{2D}=1.488$ and $k_F^2 R_s=-0.245$.
It shows that potential $V_s(r)$ has a much smaller barrier while extends to a longer range.
Our calculation shows that the breathing mode frequency $\omega_B$ are $2.078\omega$ and $2.075\omega$,
respectively, for $V(r)$ and $V_s(r)$, where the difference is around $0.15\%$.
It's worth noting that the ground state energies are $2.433E_\text{ho}$ and $2.464E_\text{ho}$, respectively, where the difference
is $1.2\%$. This shows that the beyond effective range contribution to breathing mode frequency 
is indeed negligible.

Second, we repeat the comparison for two additional data points in $(2,2)$ system.
For $k_F a_{2D}=1.488$ and $k_F^2 R_s=-0.0245$, $\omega_B$ are $2.085\omega$ and $2.080\omega$,
respectively, for $V(r)$ and $V_s(r)$.
For $k_F a_{2D}=1.120$ and $k_F^2 R_s=-0.0245$, $\omega_B$ are $2.102\omega$ and $2.095\omega$,
respectively, for $V(r)$ and $V_s(r)$. By comparing three sets of data points, 
we confirm the trend that effective range reduces quantum anomaly is
valid for a different potential. This also suggests that beyond effective range contribution to breathing mode frequency
is small to negligible.

\section{Few-body results on the density profile}

\begin{figure}[t]
\centering{}\includegraphics[bb=20bp 20bp 537bp 672bp,width=0.45\textwidth]{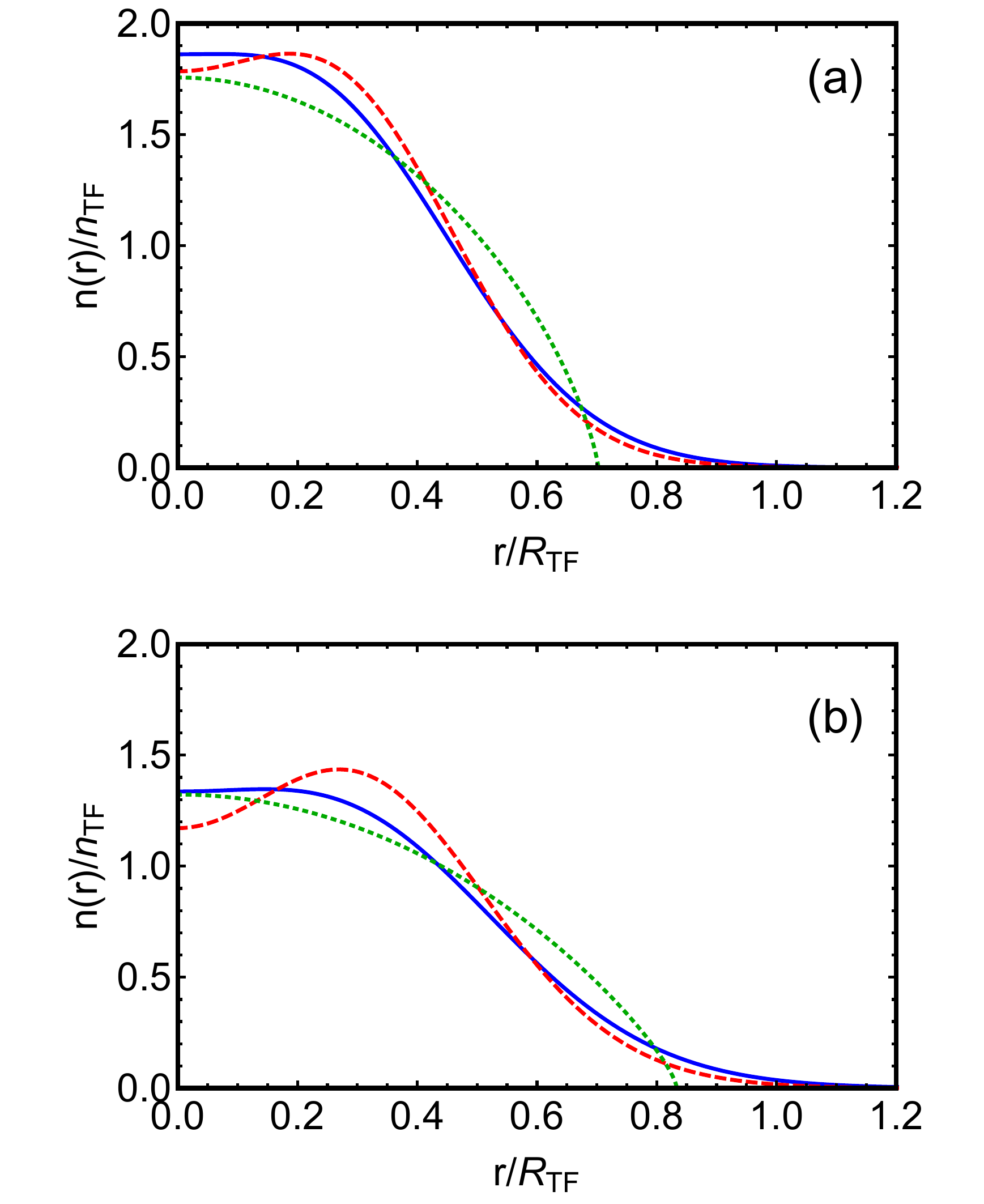}
\caption{Solid [dashed] lines in (a) and (b) show the normalized density distribution
$n(r)/n_{\text{TF}}$ of $(2,2)$ [$(3,3)$] systems for $\ln(k_{F}a_{2D})=0.101$
and $1.490$. The effective range is $k_{F}^{2}R_{s}=-0.0245$. Dotted
lines show the local density approximation results for the same $\ln(k_{F}a_{2D})$,
calculated based on the AFQMC equation of state with a zero-range
interaction \cite{shi2015}.}
\label{fig_den} 
\end{figure}

As we mentioned earlier, the agreement between our few-body results
and many-body AFQMC results for breathing mode frequency becomes worse
when $\ln(k_{F}a_{2D})<0.2$. The reason is that the Fermi wave vector
$k_{F}=\sqrt{2}(N_{\text{tot}})^{1/4}a_{\text{ho}}^{-1}$ for a trapped
two-component Fermi gas is defined at the non-interacting limit. When
$\ln(k_{F}a_{2D})$ approaches zero, the interaction becomes stronger
and deeper two-body bound state starts to form. In this regime, the
density distribution $n(r)$ becomes very different from that of non-interacting
Fermi gas $n_{0}(r)$. 

Figure~\ref{fig_den}(a) and (b) show the density distribution at
$\ln(k_{F}a_{2D})=0.101$ and $1.490$, respectively. The density
is normalized by Thomas-Fermi density $n_{\text{TF}}=(\sqrt{N_{\text{tot}}}/\pi)a_{\text{ho}}^{-2}$
while the radius is normalized by Thomas-Fermi radius $R_{\text{TF}}=(4N_{\text{tot}})^{1/4}a_{\text{ho}}$.
Solid and dashed lines show the few-body $(2,2)$ and $(3,3)$ results while the dotted lines
show the LDA results based on the AFQMC equation of state. These are
to be compared with the non-interacting density distribution $n_{0}(r)=n_{\textrm{TF}}(1-r^{2}/R_{\textrm{TF}}^{2})\Theta\left(R_{\textrm{TF}}-r\right)$.
In both few-body and LDA results, the peak density at strongly interacting
regime is much higher than the non-interacting limit, and approaches
two times the Thomas-Fermi density when $\ln(k_{F}a_{2D})$ goes to
zero. In this regime, $k_{F}$ is no longer an appropriate unit to
characterize the few-body and many-body results.

\section{Alternative extrapolation to the many-body limit}

\begin{figure}[t]
\centering{}\includegraphics[bb=20bp 20bp 537bp 672bp,width=0.5\textwidth]{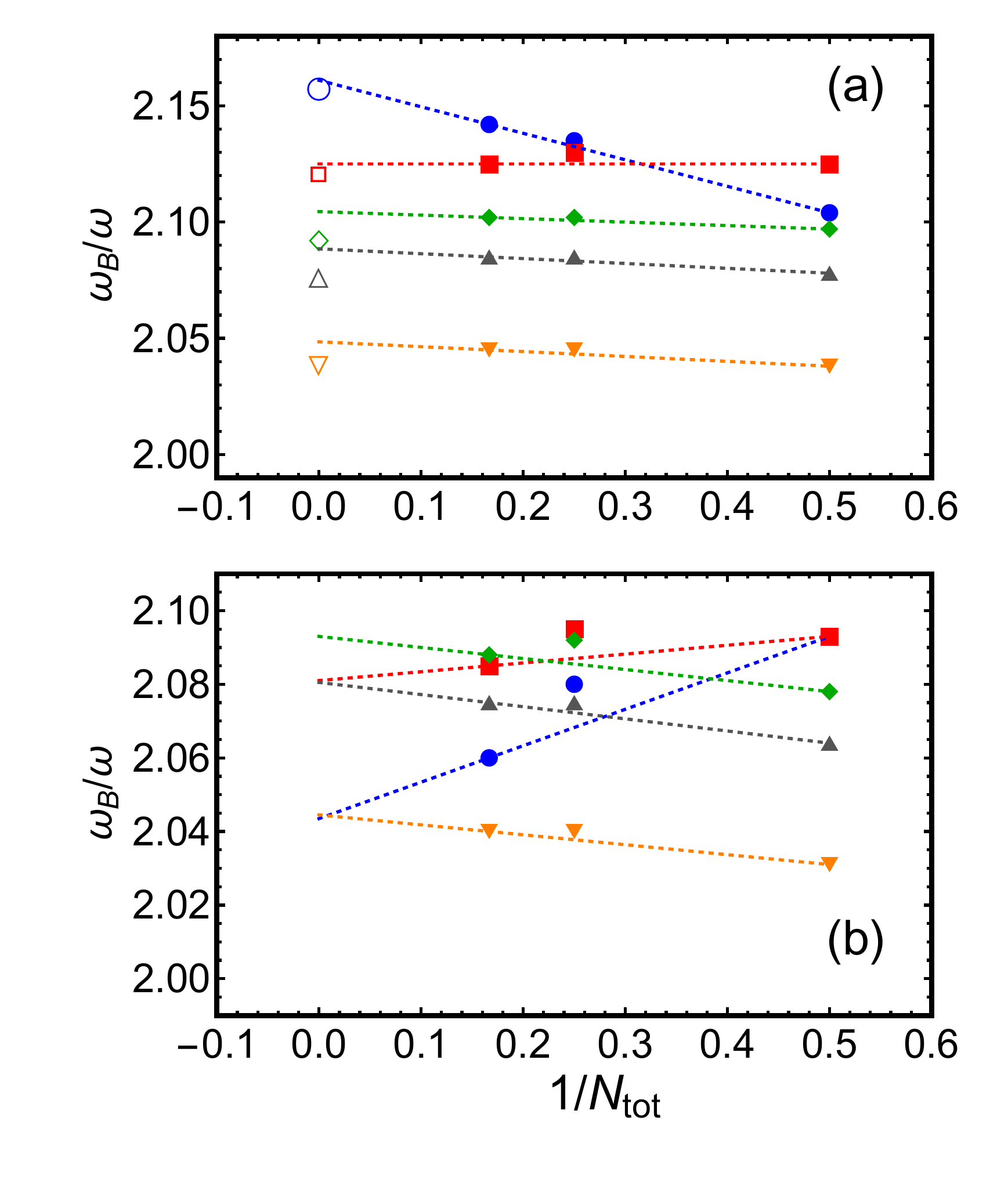}
\caption{Same as Fig.3 in the main text, except that
dotted lines show the linear extrapolation of (1,1) and (3,3) results towards the many-particle limit $N_\text{tot}\rightarrow\infty$.}
\label{fig_extra} 
\end{figure}

In the main text, we discussed a na\"ive extrapolation from few-body results to the many-body limit 
$N_\text{tot}\rightarrow\infty$. An argument can be made that both (1,1) and (3,3) systems
are closed-shell~\cite{bruun2016}. Therefore, it is worthwhile to show an alternative extrapolation 
from (1,1) and (3,3) results in Fig.~\ref{fig_extra}. The difference between the two sets of extrapolated
results are at most $1\%$.